\documentclass[a4paper]{article}
\usepackage[sort, numbers]{natbib}
\providecommand{\keywords}[1]{\textbf{\textit{Keywords: }} #1}

\usepackage[english]{babel}
\usepackage[utf8]{inputenc}
\usepackage{hyperref}
\usepackage{url}
\usepackage{amsmath}
\usepackage{graphicx}
\usepackage[titletoc,title]{appendix}
\usepackage{array}
\newcolumntype{R}[1]{>{\raggedleft\let\newline\\\arraybackslash\hspace{0pt}}m{#1}}

\title{
When a Movement Becomes a Party:
\\The 2015 Barcelona City Council Election}

\author{
Pablo Aragón, Yana Volkovich, David Laniado, Andreas Kaltenbrunner\\\\
Eurecat, Spain\\
\small\{{}name.surname\}{}@eurecat.org
}

\begin{document}
\maketitle

\begin{abstract}
Barcelona en Comú, an emerging grassroots movement-party, won the 2015 Barcelona City Council election. This candidacy was devised by activists involved in the 15M movement in order to turn citizen outrage into political change. On the one hand, the 15M movement is based on a decentralized structure. On the other hand, political science literature postulates that parties historically develop oligarchical leadership structures. This tension motivates us to examine whether Barcelona en Comú preserved a decentralizated structure or adopted a conventional centralized organization. In this article we analyse the Twitter networks of the parties that ran for this election by measuring their hierarchical structure, information efficiency and social resilience. Our results show that in Barcelona en Comú two well-defined groups co-exist: a cluster dominated by the leader and the collective accounts, and another cluster formed by the movement activists. While the former group is highly centralized like the other major parties, the latter one stands out for its decentralized, cohesive and resilient structure. 
\end{abstract}

\section{Introduction}
In the last years a new global wave of citizen protests has emerged: the Arab Spring, the 15M movement in Spain, Occupy Wall Street, \#{}YoSoy132 in Mexico, Occupy Gezi in Turkey, the Brazilian movement \#{}VemPraRua, Occupy Central in Hong Kong, etc. All these movements share common characteristics such as the claim for new models of democracy, the strategic usage of social media (e.g Twitter), and the occupation of physical spaces. 
All of them have encountered difficulties in modifying the institutional agenda and, hence, the public policies. The 2015 Barcelona City Council election is one of the first cases in which one of these movements has got to ``occupy'' the public institutions 
by building Barcelona en Comú (BeC), a political party that won the elections. BeC was conceived as the confluence of (1) minor and/or emerging parties and, to a large extent, (2) collectives and activists, with no political party affiliation, who played a prominent role in the 15M movement. 

The 15M movement, also referred to as \#{}SpanishRevolution or the ``Indignados'' Movement, emerged in May 2011 and has been defined as a ``networked social movement of the digital age''  \cite{castells2013networks}. Networked social movements, like the Arab Spring, the 15M and Occupy Wall Street, are claimed to be ``a network of networks, they can afford not to have an identifiable centre, and yet ensure coordination functions, as well as deliberation, by interaction between multiple nodes'' \cite{castells2013networks}. Other authors have formulated similar hypotheses when defining this new model of social movement as a ``change from logic of collective action, associated with high levels of organizational resources and the formation of collective identities, to a logic of connective action, based on personalized content sharing across media networks'' \cite{bennett2012logic}. We should note that some voices have refused these theoretical assumptions and argued that ``a handful of people control most of the communication flow'' and, consequently, the existence of leaders in such movements could not be denied \cite{gerbaudo2012tweets}. Empirical studies revealed that the 15M network on Twitter is characterised by its ``decentralized structure, based on coalitions of smaller organizations'' in spite of ``a small core of central users is still critical to trigger chains of messages of high orders of magnitude'' \cite{gonzalez2011dynamics}. Decentralization has been also observed in \cite{toret2015tecnopolitica} in which the 15M network is defined as open and polycentric.

The 15M network properties (i.e. 
decentralization, openness, polycentrism) could be perceived as a striking contrast to conventional political organizations, in particular, political parties. The Iron Law of Oligarchy \cite{michels1915political} postulates that political parties, like any complex organization, self-generate an elite (i.e. ``Who says organization, says oligarchy''). Although some scholars have criticised the idea that organizations will intrinsically build oligarchical leadership structures \cite{lipset1956union, rothschild1976conditions, edelstein1979comparative}, many political and social theorists have supported that, historically, small minorities hold the most power in political processes \cite{pareto1935mind, mosca1939ruling, mills1999power}. Regarding Spanish politics, a study of the 2011 national election campaign on Twitter revealed that ``minor and new parties tend to be more clustered and better connected, which implies a more cohesive community'' \cite{aragon2013communication}. Nevertheless, all the difussion networks of parties in that study were strongly centralized around their candidate and/or party profiles. Later studies analysed the interactions on Twitter between the 15M nodes and political parties and conclude that networked social movements are para-institutions: perceived as institutions but preserving an internal networked organization \cite{parainstitutions}. However, these conclusions were formulated by analysing the networks when no elections were held, before institutionalisation began. Election campaigns are competitive processes that might favor the centralization of an organization around candidates. Indeed, it has been proved that the network properties of political parties change when elections arrive \cite{Zurich}. 

Given that Barcelona en Comú emerged from the 15M and this networked movement is characterised by a decentralized structure, the research question of this study is the following: \emph{Has Barcelona en Comú preserved a decentralizated structure or has it adopted into a conventional centralized organization ruled by an elite?}

Previous hypotheses \cite{medina2015mirada} about Podemos, a member party of the Barcelona en Comú candidacy and also inspired by the 15M movement, postulate an organization formed by a \emph{front-end} (``spokesmen/spokeswomen who are visible from the media perspective'') and a \emph{back-end} (``muscle of the organization, barely visible from the media perspective''). Per contra, there are no empirical validations of this hypothesis. We strongly believe that the answer to the above research question will provide relevant insights into the institutionalisation of these new paradigm of social movements. 

Motivated by our research question, we aim to characterise the social structures of Barcelona en Comú by comparing its diffusion network on Twitter to the ones of the other political parties running for this election. The identification of the sub-network corresponding to each party is made possible by the highly divided partisan structure of the information diffusion network. This assumption relies on previous studies of political discussions in social media \cite{adamic2005political, conover2011political}. Recent research in data-driven political science has revealed the recurrent existence of boundaries between ideological online communities, in particular, political parties. A study of the 2004 U.S. Presidential election depicted a divided blogosphere in which liberals and conservatives barely generated links between the two communities \cite{adamic2005political}. Similarly, the network of retweets for the 2010 U.S. congressional midterm elections exhibited a highly segregated partisan structure where connections between left- and right-leaning users were extremely limited \cite{conover2011political}. Both studies have been taken as relevant empirical validations of the so-called \emph{cyber-balkanization}, a social phenomenon that occurs when Internet users form isolated groups around specific interests, e.g. politics. This concept is closely related to the idea of \emph{echo chambers}, in which people are ``mainly listening to louder echoes of their own voices'' \cite{sunstein2009republic} and, therefore, reinforce division in social media. Indeed, online polarization is not only a particular feature of U.S. politics but also a social behaviour observed in a diverse range of countries, e.g. Canada \cite{gruzd2014investigating} and Germany \cite{feller2011divided}. In Spain, previous studies of the Twitter networks related to recent elections also showed evidence of online polarization, e.g. in the 2010 Catalan election \cite{congosto2011twitter} and in the 2011 Spanish elections \cite{borondo2012characterising, aragon2013communication}. 

In this study, we first measure the polarization of the network, detect the online difussion sub-network of each party, and identify the users who build bridges between these clusters. Then, we analyse the diffusion networks of each of the detected clusters to characterise the social structure of the corresponding parties. The analysis of the social structures extenses the framework introduced in \cite{Zurich} which focuses on three dimensions: \emph{hierarchical structure}, \emph{effective diffusion} and \emph{social resilience}. 

\section{Data preparation}
Here, we describe the construction of the network of retweets and introduce a data-driven framework to extract the clusters corresponding to the political parties.

\subsection{Network construction}

Data were collected from Twitter in relation to the campaign for the 2015 Barcelona City Council election (May 1-26, 2015). We defined a list of Twitter accounts of the seven main political parties: 
Barcelona en Comú (BeC)\footnote{ \url{http://en.wikipedia.org/wiki/Barcelona_en_Com\%C3\%BA}}, 
Convergència i Unió (CiU)\footnote{
\url{http://en.wikipedia.org/wiki/Convergence_and_Union}}, 
Ciudadanos (Cs)\footnote{ \url{http://en.wikipedia.org/wiki/Citizens_(Spanish_political_party)}}, 
Capgirem Barcelona (CUP)\footnote{ \url{http://en.wikipedia.org/wiki/Popular_Unity_Candidates}}, 
Esquerra Republicana de Catalunya\footnote{ \url{http://en.wikipedia.org/wiki/Republican_Left_of_Catalonia}}, 
Partit Popular de Catalunya (PP)\footnote{ \url{http://en.wikipedia.org/wiki/People\%27s_Party_of_Catalonia}}, and  
Partit dels Socialistes de Catalunya (PSC)\footnote{ \url{http://en.wikipedia.org/wiki/Socialists\%27_Party_of_Catalonia}}.
We also added the Twitter accounts for corresponding candidates for Mayor and each member party for the coalitions CiU, BeC and CUP. The users of that list can be found in Table~\ref {tab:accounts}. From the Twitter Streaming API, we extracted 373,818 retweets of tweets that (1) were created by, (2) were retweeted by, or (3) mentioned a user from the list.

\begin{table}[tb!]
\centering
\advance\leftskip-0.5cm

\caption{Twitter accounts of the selected political parties and candidates.}
\label{tab:accounts}\
\begin{tabular}{@{}l|ll@{}}
Political Party / Coalition & Party account(s) &  Candidate account \\
\hline
BeC  & \begin{tabular}[c]{@{}l@{}}@bcnencomu\\ @icveuiabcn \\ @podem\_bcn\\ @equobcn \\ @pconstituentbcn\end{tabular} & @adacolau               \\
\hline
CiU  & \begin{tabular}[c]{@{}l@{}}@cdcbarcelona\\ @uniobcn\end{tabular}                                               & @xaviertrias            \\
\hline
Cs & @cs\_bcna                                                                                                      & @carinamejias           \\
\hline
CUP  & \begin{tabular}[c]{@{}l@{}}@capgirembcn \\ @cupbarcelona\end{tabular}                                          & @mjlecha    \\
\hline
ERC  & @ercbcn                                                                                                        & @alfredbosch            \\
\hline
PP  & @ppbarcelona\_                                                                                                 & @albertofdezxbcn        \\
\hline
PSC  & @pscbarcelona                                                                                                  & @jaumecollboni          
\end{tabular}
\end{table}

From this collection of retweets, we built a directed weighted graph comprising a set of nodes (users) and a set of edges (retweets between any pair of users). The weight of each edge was the number of retweets from the source node to the target node. To exclude anecdotal interactions between users and highlight the structure of the expected clusters, we only kept the interactions between any pair of nodes that occurred at least 3 times:  an edge from user A to user B implied that user A has retweeted at least three times user B in our dataset. Nodes without edges after this process were removed. The resulting network comprises 6,492 nodes and 16,775 edges.

\subsection{Community detection}
Traditionally, community detection is performed by applying a clustering algorithm. We chose the Louvain method \cite{blondel2008fast} which is commonly used because of its high performance in terms of efficiency and accuracy. Like many clustering algorithms, however, this method results into problems when defining boundaries between clusters: it assigns each node to one cluster, and also nodes that do not strongly belong to any cluster are assigned to one. The algorithm's outcome depends on the particular execution that is considered. This means that a node that appears to belong to a certain cluster could fall in another cluster if we run the algorithm another time. To solve this issue, we have designed an adapted version of the Louvain method: the algorithm is executed several times, and only the nodes that fall into the same cluster during the large majority of these executions are assigned to it. 

We first executed the standard Louvain method and found 151 clusters and achieved a modularity value of 0.727. From Figure~\ref{fig:modules-size-distribution} in Supporting Information Section we observed a clear difference between the 8 largest clusters (size $\in [232,1981]$) and the remaining 143 clusters (size $\in [2,62]$). In order to label these 8 clusters, we manually inspected the most relevant users from each cluster according to their PageRank value within the full network (the top five users for each cluster are listed in Table~\ref {tab:clusters}). The results indicate that the community detection method identified a single cluster for almost each party: BeC = $c_1,c_4$; ERC = $c_2$; CUP = $c_3$; Cs = $c_5$; CiU = $c_6$; PP = $c_7$ and PSC = $c_8$. The only exception for such rule is that BeC is composed of two clusters. The manual inspection of the users from these two clusters revealed that cluster $c_1$ is formed by the official accounts of the party (e.g. @bcnencomu, @ahorapodemos), allied parties (e.g. @ahoramadrid), the candidate (@adacolau) and a large community of peripheral users. Cluster $c_4$ is composed of activists engaged in the digital communication for the campaign (e.g. @toret, @santidemajo, @galapita). That is to say that the most visible accounts from the media perspective belong to $c_1$ while $c_4$ is formed by party activists, many of them related to the 15M movement. For this reason, from now on, we distinguish these clusters as ``{BeC-p}'' and ``{BeC-m}'', party and movement respectively. 

\begin{table}[tb!]
\centering \footnotesize
\caption{Top 5 users for the 8 largest clusters according to their PageRank value within the full network (clusters are ordered by size).}
\label{tab:clusters}
\begin{tabular}{l|l|l|l|l}
Cluster id  &Cluster label & User            & PageRank & Role          \\
\hline        &                 &          &               \\
$c_1$ & BeC-p       & @bcnencomu       & 0.092    & BeC party account \\
$c_1$ & BeC-p       & @adacolau        & 0.029    & BeC candidate     \\
$c_1$ & BeC-p       & @ahoramadrid     & 0.009    & BeC allied party account         \\
$c_1$ & BeC-p       & @ahorapodemos    & 0.009    & BeC member party account         \\
$c_1$ & BeC-p       & @elperiodico     & 0.005    & media         \\
 &        &                 &          &               \\
$c_2$ & ERC       & @ercbcn          & 0.016    & ERC party account         \\
$c_2$ & ERC       & @alfredbosch     & 0.011    & ERC candidate     \\
$c_2$ & ERC       & @naciodigital    & 0.009    & media         \\
$c_2$ & ERC       & @arapolitica     & 0.007    & media         \\
$c_2$ & ERC       & @esquerra\_erc   & 0.004    & ERC party account         \\
 &        &                 &          &               \\
$c_3$ & CUP      & @cupbarcelona    & 0.016    & CUP party account         \\
$c_3$ & CUP      & @capgirembcn     & 0.008    & CUP party account         \\
$c_3$ & CUP      & @albertmartnez   & 0.005    & media         \\
$c_3$ & CUP      & @encampanya      & 0.003    & media         \\
$c_3$ & CUP      & @mjlecha         & 0.002    & CUP candidate     \\
 &        &                 &          &               \\
$c_4$ & BeC-m       & @toret           & 0.014    & BeC member  \\
$c_4$ & BeC-m       & @santidemajo     & 0.005    & BeC member  \\
$c_4$ & BeC-m       & @sentitcritic    & 0.005    & media         \\
$c_4$ & BeC-m       & @galapita        & 0.005    & BeC member  \\
$c_4$ & BeC-m       & @eloibadia       & 0.005    & BeC member  \\
 &        &                 &          &               \\
$c_5$ & Cs      & @carinamejias    & 0.007    & Cs candidate     \\
$c_5$ & Cs      & @cs\_bcna        & 0.006    & Cs party account         \\
$c_5$ & Cs      & @ciudadanoscs    & 0.004    & Cs party account         \\
$c_5$ & Cs      & @soniasi02       & 0.003    & Cs member  \\
$c_5$ & Cs      & @prensacs        & 0.002    & media         \\
 &        &                 &          &               \\
$c_6$ & CiU      & @xaviertrias     & 0.012    & CiU candidate     \\
$c_6$ & CiU      & @ciu             & 0.004    & CiU party account         \\
$c_6$ & CiU      & @bcn\_ajuntament & 0.003    & institutional \\
$c_6$ & CiU      & @ramontremosa    & 0.002    & CiU member  \\
$c_6$ & CiU      & @cdcbarcelona    & 0.002    & CiU party account     \\
 &        &                 &          &               \\
$c_7$ & PP     & @btvnoticies     & 0.011    & media         \\
$c_7$ & PP      & @cati\_bcn       & 0.003    & media         \\
$c_7$ & PP      & @albertofdezxbcn & 0.003    & PP candidate     \\
$c_7$ & PP      & @maticatradio    & 0.002    & media         \\
$c_7$ & PP      & @ppbarcelona\_   & 0.002    & PP party account     \\   
 &        &                 &          &               \\
$c_8$ & PSC      & @elsmatins       & 0.006    & media         \\
$c_8$ & PSC      & @pscbarcelona    & 0.003    & PSC party account         \\
$c_8$ & PSC      & @sergifor        & 0.003    & media         \\
$c_8$ & PSC      & @jaumecollboni   & 0.002    & PSC candidate     \\
$c_8$ & PSC      & @elpaiscat       & 0.002    & media         \\
\end{tabular}
\end{table}

Furthermore, we found a remarkable presence of accounts related  to media in Table~\ref {tab:clusters} for almost every cluster. As we noted above, we aim to study the ecosystem of each political party, i.e including only nodes that are reliably assigned to them. To this end, we applied the adapted version of the Louvain method that is described in Section~\ref{sub:robust}: we ran the algorithm 100 times and assigned to each cluster only the nodes that fell into that cluster more than 95 times. By inspecting the results of the 100 executions, we found the presence of 8 major clusters, much bigger than the others, as a constant element. The composition of these clusters is also quite stable: 4,973 nodes (82.25\%) are assigned to the same cluster in over 95 executions. 

Among the remaining nodes, which could not be reliably assigned to any of the major clusters, we find that many accounts are media. We additionally identified the most relevant users, according to PageRank, in the sub-network formed only by edges between nodes from different clusters (i.e. “weak ties” \cite{granovetter1973strength}). Table~\ref {table:weakties} presents the 25 most relevant users in this sub-network and confirms that media played a key role in connecting different clusters.

\section{Results}
So far we have described the way diffussion network was constructed and the ways it was divided into cluster corresponding to major political parties. For the next steps we focus on polarization of the network during the campaign and compare structural peculiarities of the largest clusters in the following dimensions: hierarchical structure, effective difussion, and social resilience.

\newpage\subsection{Polarization} 
\begin{figure}[htbp]
\centering
\advance\leftskip-4cm
\includegraphics[width=1.5\textwidth]{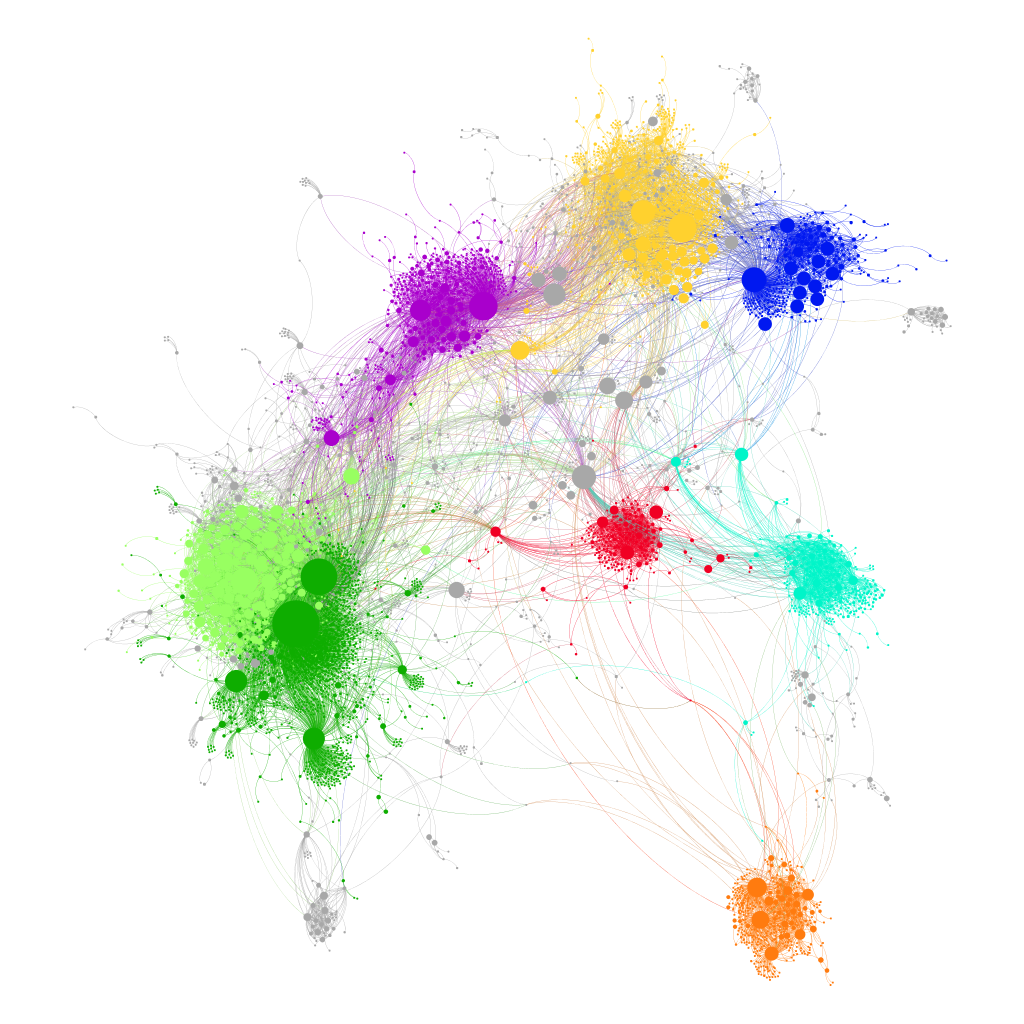}
\caption{\label{fig:bcn2015} Network of retweets (giant component). Clusters are represented by color: {BeC-p} in dark green; {BeC-m} in light green; ERCin yellow; PSC in red; CUP in violet; Cs in orange; CiU in dark blue; PP in cyan. The nodes out of these clusters are grey-colored.}
\end{figure}

Similar to the previous findings for online political networks we detected a high level of polarization when calculated modularity ($Q=0.727$) for the first execution of standard Louvain method as described before. The boundaries between ideological online communities are visible in Figure~\ref{fig:bcn2015}, where we visualized the resulting graph partitioning for $N=100$ and $\varepsilon=0.05$. For a better readability of the network, we only considered the giant component of the graph and applied the Force Atlas 2 layout algorithm \cite{jacomy2011force} to enforce cluster graph drawing.

\begin{figure}[htbp]
\centering
\includegraphics[width=0.9\textwidth]{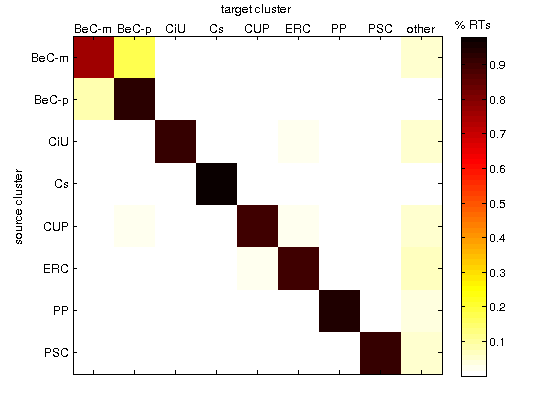}
\caption{\label{fig:adjacency} Normalized weighted adjacency matrix of the network of clusters.}
\end{figure}

As one could expected in any polarized scenario, the largest number of interaction links happened within the same cluster. There were however a notably large number of links between the two clusters of BeC ({BeC-p} and {BeC-m}). To further prove the low levels of interactions between major parties we made an interaction matrix $A$, where $A_{i,j}$ counts all retweets that accounts assigned to cluster $i$ made for the tweets from users of cluster $j$. Since the clusters are of the different size, we then normalized $A_{i,j}$'s by the sum of the all retweets made by the users assigned to cluster $i$. From Figure~\ref{fig:adjacency}, where we draw matrix $A$, we confirmed that a vast majority of retweets were made between users from the same cluster (main diagonal). For Barcelona en Com\'u we found a presence of communication between movement and party clusters with a prevalence from {BeC-m} to {BeC-p} (0.18, the largest value out of the main diagonal). 

\subsection{Structure of the party clusters}
Inspired by the framework introduced in \cite{Zurich} we proposed to compare the topology of the intra-network of each cluster in terms of hierarchical structure, information efficiency, and social resilience. 

\subsubsection{Hierarchical structure}
To evaluate the hierarchical structure we measured the in-degree inequality of each cluster based on the Gini coefficient. We also calculated in-degree centralization suggested in~\cite{Zurich}, however found it uninformative in the case of high variability of the data. 

\begin{figure}[htbp]
\centering
\advance\leftskip-1cm
\includegraphics[width=1\textwidth]{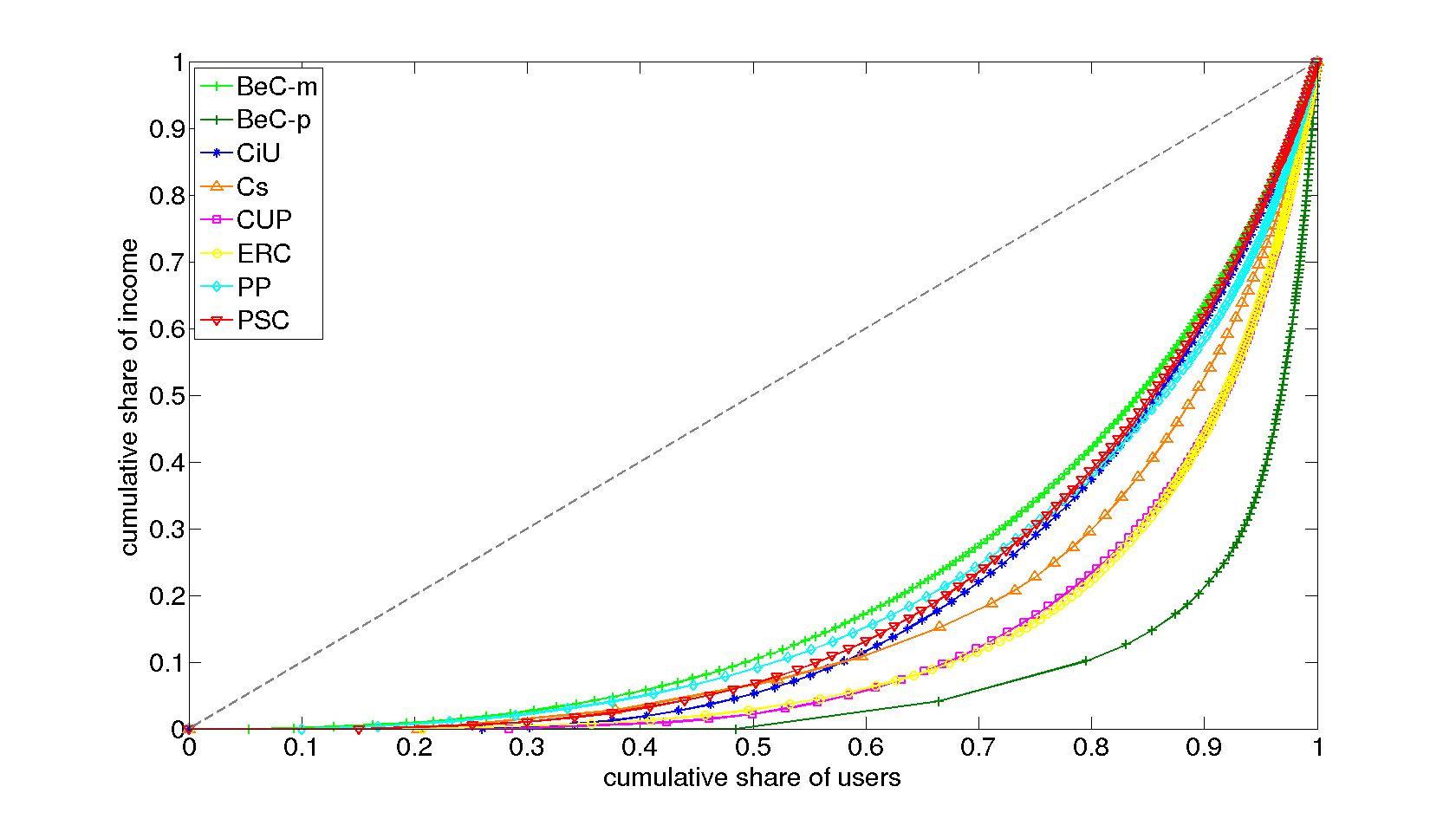}
\caption{\label{fig:in-degree} Lorenz curve of the in-degree distribution of each cluster.}
\end{figure}

From results in Table~\ref{gini} we saw a notable divergence between these hierarchical metrics: the inequality values of CiU and PP are similar ($G_{in}=0.893$ and $G_{in}=0.876$, respectively), but PP centralization ($C_{in}=0.378$) is far from the maximum centralization value exhibited by CiU ($C_{in}=0.770$).  For Barcelona en Com\'u, {BeC-m} emerges as the least inequal and the least centralized structure, while {BeC-p} forms the most inequal cluster ($G_{in}=0.995$). We also plotted the Lorenz curve of the in-degree distribution of the clusters in Figure~\ref{fig:in-degree} to visually validate the different levels of inequality among clusters that were presented in Table~\ref{gini}.

It is easy to demonstrate that for networks with a heavy tailed in-degree distribution (as the ones of this study) the in-degree centralization formulated in \cite{freeman1979centrality} is approximately equal to the ratio between the maximum in-degree and the number of nodes\footnote{ This is caused by the differences of several orders of magnitude between the maximum and average in-degree, common situation for social graphs.}. Therefore, this metric is not a good one to capture hierarchical structure for social diffusion graphs, and Gini coefficient for in-degree inequality represents a more reliable measure.

\begin{table}[tb!]
\centering
\caption{Inequality based on the Gini Coefficient ($G_{in}$) and centralization ($C_{in}$) of the in-degree distribution of each cluster.}
\label{gini}
\begin{tabular}{l|ll}
Cluster & $G_{in}$ & $C_{in}$ \\
\hline
BeC-p & 0.995 & 0.639 \\
Cs   & 0.964 & 0.476 \\
ERC  & 0.954 & 0.452 \\
CUP  & 0.953 & 0.635 \\
CiU  & 0.893 & 0.770 \\
PP   & 0.876 & 0.378 \\
PSC  & 0.818 & 0.565 \\
BeC-m & 0.811 & 0.290 \\
\end{tabular}
\end{table}

\subsubsection{Information efficiency}
Broadly speaking the efficiency of a network aims to measure its small-world property, i.e.   phenomenon of strangers being linked by a mutual acquaintance. To assess the efficiency of information transportation within each party cluster we computed the average path length and the clustering coefficient. Small-world networks tend to have a small average shortest path length and a clustering coefficient significantly higher than expected by random chance \cite{watts1998collective}. 

From Table~\ref {tab:cluster-distance} we observe that {BeC-m} has the highest clustering coefficient ($Cl=0.208$) closely followed by PP and PSC, the two smallest clusters by size. On the contrary the clustering coefficient of {BeC-p} is almost 0. This finding could be explained by the topology of {BeC-p}, roughly formed by stars whose center nodes are the most visible Twitter accounts of Barcelona en Comú: the party official accounts and the candidate. 

We do not observe a remarkable pattern regarding the average path length. It is lower than 3 for the majority of the party clusters with the PSC cluster having the lowest value ($l=2.29$). In the same time ERC, CiU and {BeC-m} expose the longest average path length (5.43, 4.66, 3.35 respectively).

\begin{table}[tb!]
\centering
\caption{Number of nodes (N) and edges (E), clustering coefficient (Cl) and average path length (l) of the intra-network of each cluster.}
\label{tab:cluster-distance}
\begin{tabular}{l|rrrr}
Cluster & N & E & Cl & l \\
\hline
{BeC-m}    & 427   & 2 431  & 0.208 & 3.35 \\
PP      & 301   & 1 163  & 0.188 & 2.73 \\
PSC     & 211   & 810    & 0.182 & 2.29 \\
CiU     & 337   & 1 003  & 0.114 & 4.66 \\
Cs      & 352   & 832    & 0.073 & 2.57 \\
CUP     & 635   & 1 422  & 0.037 & 2.57 \\
ERC     & 866   & 1 899  & 0.027 & 5.43 \\
{BeC-p}    & 1 844 & 2 427  & 0.002 & 2.48           
\end{tabular}
\end{table}

\subsubsection{Social resilience}
The concept of social resilience is the ability of a social group to withstand external stresses. To measure social resilience for a social network we applied the k-core decomposition for each cluster and evaluated the distributions of the nodes within each k-core. 

\begin{figure}[htbp]
\centering
\includegraphics[width=0.9\textwidth]{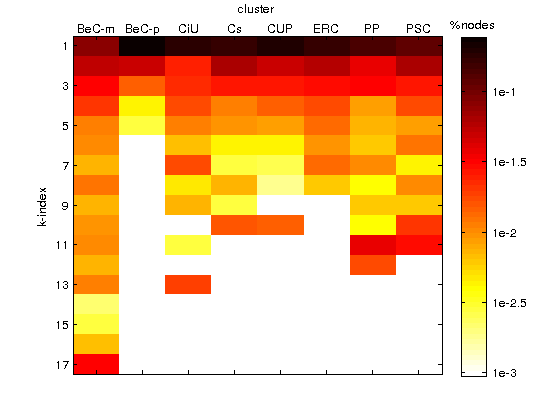}
\caption{\label{fig:kmatrix} Distribution of the nodes per cluster (column) and k-index (row). Cells are colored to form a heat map indicating the density (log scale).}
\end{figure}

In Table~\ref {tab:kcore} we present maximal and average k-indexes for each cluster and Figure~\ref{fig:kmatrix} visually shows the corresponding distributions. As in the case of hierarchical structure and information efficiency we observed a remarkable difference between {BeC-m} ($k_{max}=17$, $k_{avg}=5.90$) and {BeC-p} ($k_{max}=5$, $k_{avg}=1.33$), that are the highest and lowest values respectively. In comparison to the other parties we saw clear differences between node distributions for both, {BeC-m} and {BeC-p}, and the rest (the largest concentration of the nodes is in the first k-cores and considerable part is in the inner most cores). Therefore, the movement group of Barcelona en Com\'u is an online social community with an extreme ability to withstand or recover. In the same time the party group of  Barcelona en Com\'u seems to only focus on the core users. 

\begin{table}[tb!]
\centering
\caption{Maximal and average k-index (standard deviation in parentheses) for the intra-network of each cluster.}
\label{tab:kcore}
\begin{tabular}{l|R{1cm} R{2cm}}
cluster & $k_{max}$ & $k_{avg}$\\
\hline
BeC-m  & 17    & 5.90  (5.46)    \\
PP      & 12    & 4.02  (3.99)    \\
PSC     & 11    & 3.85  (3.55)    \\
CiU     & 13    & 3.10  (3.44)    \\
ERC     & 8     & 2.25  (1.85)    \\
Cs      & 10    & 2.42  (2.42)    \\
CUP     & 10    & 2.19  (2.22)    \\
BeC-p   & 5     & 1.33  (0.71)   
\end{tabular}
\end{table}

\section{Discussion}

In this section, we discuss the results from examining the structures on Twitter of the political parties in the 2015 Barcelona City Council election. 

\subsection{Institutionalisation of a networked movement}

Our research question deals with the kind of organizational structure that Barcelona en Comú developed for the campaign. On the one hand, the cited literature \cite{gonzalez2011dynamics, toret2015tecnopolitica} provided evidence of the decentralization of the 15M movement, which inspired the Barcelona en Comú candidacy. On the other hand, many political scientists \cite{michels1915political, pareto1935mind, mosca1939ruling, mills1999power} argued that parties are always 
ruled by elites and, therefore, result in centralized organizations. Furthermore, the historical models of political parties reviewed in \cite{katz1995changing} (i.e. \emph{Caucus parties}, \emph{Mass parties}, \emph{Catch-all parties}, and \emph{Cartel parties}) always assumed organization around elites. All of these observations motivated us to study whether Barcelona en Comú preserved a decentralizated structure or adopted a conventional centralized organization.

Our results depict a movement-parrty structure in which the two components form well-defined clusters. In comparison to the clusters of the rest of political parties, we found the BeC movement community as the least hierarchichal, better clustered and most resilient one. In contrast, the BeC party community emerges as the most hierarchical, least clustered and least resilient one. The centralization of the party cluster points to the candidate and official accounts, the subjects that are commonly associated with the elite. However, unlike the rest of political parties, there is a co-existence of both party and movement clusters. This co-existence is consistent with the hypothesis expressed in \cite{medina2015mirada} when defining Podemos, member party of Barcelona en Comú, as the conjugation of a front-end and a back-end. 

In this article we have characterised the organization of political parties according to their online diffusion networks. Some authors have reported that the Internet played a key role in the organization of the 15M movement for building ``a hybrid space between the Internet social networks and the occupied urban space'' \cite{castells2013networks}. According to \cite{toret2015tecnopolitica}, this hybrid space is the result of \emph{technopolitical practices}: ``the tactical and strategic use of technological devices (including social networks) for organization, communication and collective action''. Are technopolitics the origin of this particular movement-party partition of Barcelona en Comú? Recently, political scientists have postulated the emergence of \emph{cyber parties} ``with its origins in developments in media and information and communication technologies'' \cite{margetts2001cyber}. Although we can not ensure that the Internet is the only reason behind this new form of political organization, in this particular context some party activists reported that ICT technologies becomes essential for campaigning \cite{lali}. Therefore, we are convinced of the close link between technopolitics and the structure of Barcelona en Comú.

\subsection{Online polarization}

The identification of the different clusters was made possible by the high level of polarization that the network exhibited, as we initially expected. We observed that bridges between clusters (i.e. ``weak ties" \cite{granovetter1973strength}) were mostly built by accounts related to media. Because media accounts hardly retweet content from other accounts, a great amount of weak ties consists of users from political clusters retweeting content published by media accounts. This means that media play a key role in generating messages that build a public sphere. Some theorists suggest that the best response to group polarization is the usage of  ``mechanisms providing a public sphere'' \cite{sunstein1999law}. We found that the most relevant account in the sub-network of weak ties was @btvnoticies, the local and publicly owned television. Indeed, this TV channel organized the debate among the candidates of five of the seven parties. Figure~\ref{fig:weakties} presents the ego-networks of four media accounts: @btvnoticies, @arapolitica @elpaiscat and @naciodigital. We clearly observe that @btvnoticies is linked from every party while the other 3 private media are only linked from specific clusters. This finding might indicate that public TV became more plural than the other three analysed private media, and pluralism is an effective tool to get ``people exposed to a range of reasonable competing views'' \cite{sunstein1999law}. 

\section{Conclusions}
In this study we have examined the Twitter networks of Barcelona en Comú in comparison to the other parties for the 2015 Barcelona municipal elections. We observed that the tension between the decentralization of networked movements and the centralization of traditional political parties results into a movement-party structure:  the two paradigms co-exist in two well-defined clusters. From this result, we find of interest to further investigate the origin of this particular structure: (1) Did the structure of Barcelona en Comú result from the confluence of minor parties and the 15M activists? Or (2) instead of evolving into a centralized organization, did the 15M networked movement implement a party interface over its decentralized structure? Further longitudinal analyses of the formation of these networks could help us to provide answer to these open questions.

It is interesting to note that city council elections were held in every Spanish city in May 2015 and candidacies similar to Barcelona en Comú were built. Moreover, after these elections, the city councils of several of the largest Spanish cities are ruled by these new organizations (e.g. Ahora Madrid, Zaragoza en Común). For this reason, future work should replicate this analysis to examine whether the characteristics that we observed in Barcelona en Comú are also present in these other grassroots movement-parties. 


\section*{Acknowledgments}

This research is supported by the EU project D-CENT (FP7/CAPS 610349). The data of this work were collected with KALIUM, a R+D project funded by the ACC1\'0 program (Generalitat de Catalunya). We would like to thank DatAnalysis15M Research Network and the \#{}GlobalRevExp Forum for their valuable discussions and suggestions that helped to improve this study. Yana Volkovich is supported by the People Programme (Marie Curie Actions, from the FP7/2007-2013) under grant agreement no.600388 managed by REA and ACC1\'O.

\newpage
\setcitestyle{square}
\bibliographystyle{ieeetr}
\bibliography{references}

\begin{appendices}

\renewcommand\appendixname{}

\renewcommand{\thesection}{S\arabic{section}}   
\renewcommand{\thetable}{S\arabic{table}}   
\renewcommand{\thefigure}{S\arabic{figure}}
\setcounter{table}{0}
\setcounter{figure}{0}

\newpage\section{Supporting Information}

The supportive figures and tables of this article are presented below.

\begin{figure}[htbp]
\centering
\includegraphics[width=1\textwidth]{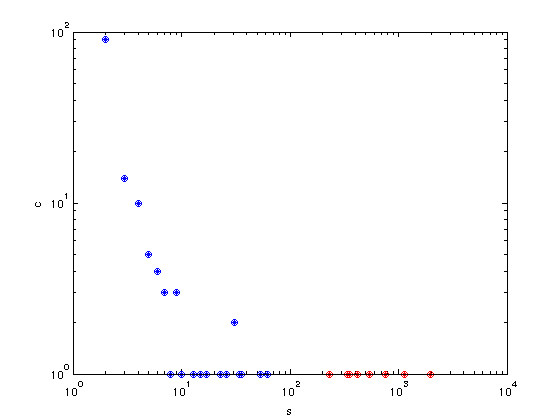}
\caption{\label{fig:modules-size-distribution} Distribution of the number of clusters (c) by size (s). Red markers are used to indicate the 8 largest clusters.}
\end{figure}

\begin{table}[tb!]
\centering
\caption{Most relevant nodes by PageRank in the sub-network formed by edges between nodes from different clusters.}
\label{table:weakties}
\begin{tabular}{l|ll}
User              & Page Rank & Role      \\
\hline
@btvnoticies     & 0.014    & media        \\
@bcnencomu       & 0.012    & party        \\
@sicomtelevision & 0.010    & media        \\
@cupbarcelona    & 0.007    & party        \\
@elsmatins       & 0.007    & media        \\
@capgirembcn     & 0.006    & party        \\
@tv3cat          & 0.006    & media        \\
@324cat          & 0.006    & media        \\
@xaviertrias     & 0.005    & candidate    \\
@puntcattv3      & 0.005    & media        \\
@revolucio1984   & 0.004    & citizen      \\
@sergifor        & 0.004    & media        \\
@nuriapujadas    & 0.004    & media        \\
@annatorrasfont  & 0.004    & media        \\
@arapolitica     & 0.004    & media        \\
@maticatradio    & 0.003    & media        \\
@cati\_bcn       & 0.003    & media        \\
@elpaiscat       & 0.003    & media        \\
@encampanya      & 0.003    & media        \\
@albertmartnez   & 0.002    & media        \\
@naciodigital    & 0.002    & media        \\
@adacolau        & 0.002    & candidate    \\
@ramontremosa    & 0.002    & party member \\
@alfredbosch     & 0.002    & candidate    \\
@directe         & 0.001    & media       
\end{tabular}
\end{table}

\begin{figure}[htbp]

\centering
\advance\leftskip-2.5cm
\includegraphics[width=1.5\textwidth]{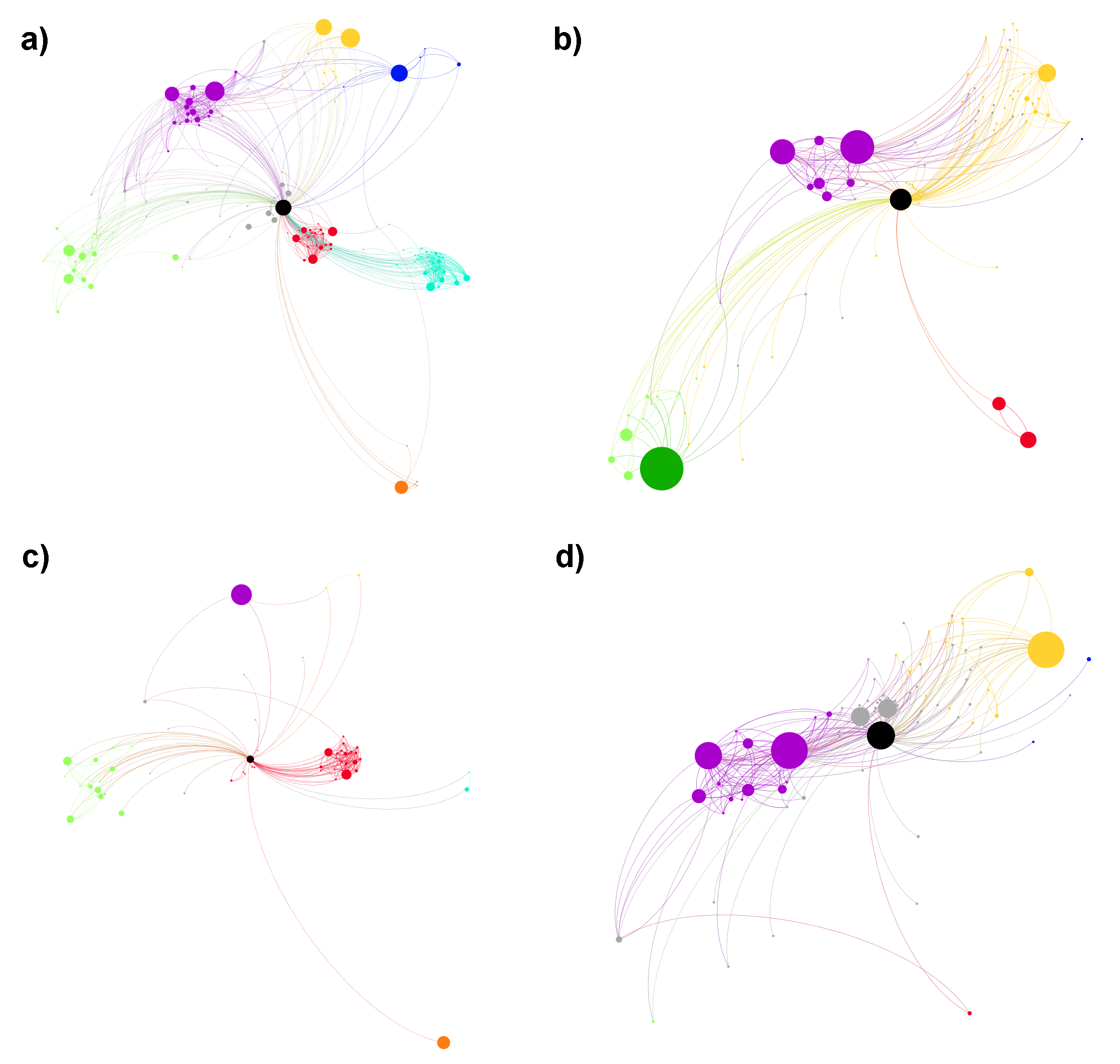}
\caption{\label{fig:weakties} Ego-networks of 4 media accounts: a) @btvnoticies; b) @arapolitica; c) @elpaiscat ; d) @naciodigital. Central nodes (i.e. corresponding media accounts) are black-colored.}
\end{figure}

\newpage\section{Methods}\label{sec:methods}
In this study we have used methods of social network analysis for community detection, identification of relevant nodes, and measurement of the topological structure of a network.

\subsection{Community detection}

\subsubsection{Modularity}
The modularity measures the density of edges inside communities in comparison to edges between communities \cite{newman2004analysis}. Its value, between -1 and 1, is defined as:
\[
Q = \frac{1}{2m}\Sigma_{ij}\bigg[A_{ij} - \frac{k_i k_j}{2m}\bigg]\delta (c_i,c_j).
\]
Here $A_{ij}$ is the edge weight between nodes $i$ and $j$; $k_i$ and $k_j$ are the degrees of the nodes $i$ and $j$, respectively; $m$ represents the total number of edges in the graph. $c_i$; and $c_j$ are the communities of the nodes and $\delta$ is a simple delta function.

\subsubsection{Louvain Method}\label{sub:robust}

The Louvain Method is a community detection technique based on a greedy algorithm that attempts to optimize the modularity of a partition of a given network \cite{blondel2008fast}. The method follows a two-step approach. 

First, each node is assigned to its own community. Then, for each node $i$, the change in modularity is measured for moving $i$ from its own community into the community of each neighbor $j$:
\[
 \Delta Q = \bigg[ \frac{\Sigma_{in} + k_{i,in}}{2m} - \bigg(\frac{\Sigma_{tot} + k_i}{2m}\bigg)^2 \bigg]-\bigg[\frac{\Sigma_{in}}{2m} - \bigg(\frac{\Sigma_{tot}}{2m}\bigg)^2-\bigg(\frac{k_i}{2m}\bigg)^2\bigg] ,
\]
where $\Sigma_{in}$ is sum of all the weights of the intra-edges of the community where $i$ being moved into, $\Sigma_{tot}$ is the sum of all the weights of the edges to nodes of the community, $k_i$ is the degree of $i$, $k_{i,in}$ is the sum of the weights of the edges between $i$ and other nodes in the community, and $m$ is the sum of the weights of all edges in the network. Once this value is measured for all communities that $i$ is connected to, the algorithm locates $i$ into the community that produces the largest increase in modularity. If no increase is possible, $i$ remains in its original community. This process is applied iteratively until modularity can not be increase and a local maximum of modularity is achieved.

In the second step, the method groups all of the nodes from the same community and builds a new network where nodes are the communities from the previous step. Edges between nodes of the same community are represented by self-loops and edges from multiple nodes from the same community to a node of a different community are represented by weighted edges between corresponding communities. 

First and second steps are repeated until modularity can not be increased.

\paragraph{Adapted version to enhance the robustness of the largest clusters}
Like most community detection methods, the Louvain method consists of a greedy algorithm and has a random component, so each execution produces a different result. To obtain robust results, avoiding dependency on a particular execution of the algorithm, we introduce the following method to identify the main clusters of the network in a stable way.

First, we run $N$ executions of the Louvain algorithm, which produce $N$ different partitions of the network into clusters.
Then we select the bigger clusters for each partition, and identify each cluster through its most representative nodes. In particular, as we expect that the main clusters will represent the political parties, we identify each cluster with the most central node corresponding to the account of a political party or of a political party leader. 
Finally, we assign to each cluster all the nodes that appear in that cluster in at least the  $1-\varepsilon$
of the partitions created, where $1-\varepsilon$ represents the confidence interval.

This procedure allows us to validate the results of the community detection algorithm, and to guarantee that all the nodes that are assigned to a cluster do actually belong to it with high confidence. 
The remaining nodes, that cannot be assigned in a stable way to any of the main clusters, are left out from all the clusters.

\subsection{Identification of relevant nodes: PageRank}

PageRank is a global characteristic of a node participation in some network and could be seen as a characteristic of node’s success and popularity \cite{Brin1998}. It is defined as a stationary distribution of a random walk on the directed graph. At each step, with probability $c$, the random walk follows a randomly chosen outgoing edge from a node, and with probability $(1-c)$ the walk starts afresh from a node chosen uniformly among all nodes. The constant c is called damping factor, and takes values between 0 and 1 (traditionally $c$=0.85). PageRank can be summarized in the following formula:

\[
PR(i) = c \sum_{j \rightarrow i} \frac{1}{d_j} PR(j) + \frac{1-c}{n},
\]

where $PR(i)$ is the PageRank of node $i$, $d_j$ is out-degree of node $j$, the sum is taken over all nodes $j$ that link to node $i$, and $n$ is the number of nodes in the network. Unlike in- and out-degree which are local characteristics, the PageRank is a global characteristic of a node. In other words, adding/removing an edge between two nodes could affect PageRank values of many nodes. 

\subsection{Network topology}

\subsubsection{In-degree distribution}
The in-degree of node $i$ is the total number of edges onto node $i$. By counting how many nodes have each in-degree value, the in-degree distribution $P(k_{in})$ is equal to the fraction of nodes in the graph with such in-degree $k_{in}$. The cumulative in-degree distribution $P(K \ge k_{in})$ represents the fraction of nodes in the graph whose in-degree is greater than or equal to $k_{in}$.

\subsubsection{In-degree Centralization}

A existing method to measure degree centralization was introduced by \cite{freeman1979centrality} and is based on two concepts: (1) how the centrality of the most central node exceeds the centrality of all other nodes and (2) setting the value as a ratio by comparing to a star network:
\[
C_{in} = \frac{\sum\limits_{i=1}^n [k^{in}_{*} - k^{in}_{i}]}{ \max \sum\limits_{i=1}^n [k^{in}_{*} - k^{in}_{i}] } ,
\]

where $k^{in}_{i}$ is the in-degree of node $i$, $k^{in}_{*}$ is the maximum in-degree of the network and $\max \sum\limits_{i=1}^n [k^{in}_{*} - k^{in}_{i}]$ is the maximum possible sum of differences for a graph with the same number of nodes (a star network).

\subsubsection{In-degree Inequality: Gini coefficient}

The Gini coefficient is a statistical metric to quantify the level of inequality given a distribution \cite{gini1912variabilita}. It was initially formulated in Economics to measure the income distribution by using the Lorenz curve. If $A$ is the area between the line corresponding perfect equality and $B$ is the area under the Lorenz curve, the Gini coefficient is equal to $A/(A+B)$. If the Lorenz curve is expressed by the function $Y = L(X)$, $B$ is calculated as follows:
\[
G = 1 -2 \int_0^1 L(X)dX
\]
In the context of network topology, the Gini coefficient can be applied to characterize the hierarchical structure of a network based on the inequality of its in-degree distribution. 

\subsubsection{Clustering coefficient}
Clustering coefficient measures the extent of nodes to clusted together by calculating the number of triangles in the network. For every node $i$ we set $N_i$ to be the neighborhood, i.e. $N_j=\{j\in V: (i,j)\in E\}$, and define the local clustering coefficient as 
\[Cl_i=\frac{2|(j,k)\in E: j,\;k\in N_j|}{k_i(k_i-1)}.\]
Then, following~\cite{watts1998collective} the clustering coefficient is just the average of the local clustering coefficients: $Cl=\sum_{i}Cl_i/n,$ where 
$n$ is the number of nodes in the network.

\subsubsection{Average path length}
The concept of average path length aims to measure the efficiency of information propagation in a social network by taking the mean value of the number of edges  along the shortest paths for all possible pairs of nodes. In more details, for every pair of nodes $i,j$ we set $d_{ij}$ to be the smallest number of steps among all directed paths between $i$ and $j$ and $d_{ij}=0$ if there is no such path. Then, the average path length is defined as follows: $l={\sum_{i\neq j}d_{ij}}/{n(n-1)}$.

\subsubsection{k-core decomposition}
The k-core of a graph is the maximal subgraph in which each vertex is adjacent, ignoring the direction of the edge, to at least k other nodes of the subgraph. A graph's node has a k-index equals to k if it belongs to the k-core but not to the (k+1)-core. Thus, a given network, we define a sub-network H induced by the subset of users C. H is a k-core of the network if and only if for every user in C: $deg_H(i) \leq k$ , and H is the maximum sub-graph which fulfils this condition. With $deg_H(i)$ we denote the degree of the node i in the sub-graph H. A user has k-index equal k if it belongs to the k-core but not to the (k+1)-core. 

In simple words, k-core decomposition starts with k = 1 and removes all nodes with degree equal to 1. The procedure is repeated iteratively until no vertices with degree 1 remain. Next, all removed nodes are assigned k-index to be 1. It continues with the same procedure for k = 2 and obtains vertices with indexes equal 2, and so on. The process stops when the last node from the network is removed at the ${k_{max}}^{th}$ step. The variable $k_{max}$ is then the maximum shell index of the graph.

\end{appendices}

\end{document}